\documentclass[aps,prc,twocolumn,floatfix,superscriptaddress,nofootinbib]{revtex4-2}
\usepackage{graphicx}
\usepackage[pdftex,colorlinks,citecolor=blue,bookmarks]{hyperref}
\usepackage{bm}
\usepackage{amsmath}
\usepackage{longtable}

\begin{document}
% \draft
\title{Mean-field proton-neutron pairing correlations with the Gogny D1S energy density functional}
\author{Miguel de la Fuente} 
\affiliation{Departamento de F\'isica Te\'orica, Universidad Aut\'onoma de Madrid, E-28049 Madrid, Spain}
\affiliation{Centro de Investigaci\'on Avanzada en F\'isica Fundamental-CIAFF-UAM, E-28049 Madrid, Spain}
\author{Tom\'as R. Rodr\'iguez} 
\affiliation{Departamento de F\'isica At\'omica, Molecular y Nuclear, Universidad de Sevilla, E-41012 Sevilla, Spain}
\author{Luis M. Robledo} 
\affiliation{Departamento de F\'isica Te\'orica, Universidad Aut\'onoma de Madrid, E-28049 Madrid, Spain}
\affiliation{Centro de Investigaci\'on Avanzada en F\'isica Fundamental-CIAFF-UAM, E-28049 Madrid, Spain}
\author{Benjamin Bally} 
\affiliation{Technische Universit\"at Darmstadt, Department of Physics, 64289 Darmstadt, Germany}
\author{Nathalie Pillet} 
\affiliation{CEA, DAM, DIF, F-91297 Arpajon, France}
\affiliation{Universit\'e Paris-Saclay, CEA, LMCE, 91680 Bruy\`eres-le-Ch\^atel, France}
\begin{abstract}
%We study proton-neutron pairing correlations within the Hartree-Fock-Bogoliubov (HFB) framework using Gogny-type energy density functionals. By allowing for proton-neutron mixing in the quasi-particle transformation, both isovector ($T=1$) and isoscalar ($T=0$) pairing channels are explicitly included at the mean-field level. The \texttt{TAURUS} code has been extended to treat density-dependent Gogny interactions in this generalized HFB scheme. We examine the numerical behavior of the widely used Gogny D1S functional and compare it with calculations performed using the Hamiltonian-based Brink-Boecker B1 interaction supplemented by a zero-range spin-orbit term. When proton-neutron mixing is included and large single-particle spaces are employed, instabilities are observed for Gogny D1S due to the zero-range density-dependent term contribution to the proton-neutron pairing field, whereas stable solutions are obtained with the B1 interaction. Constrained HFB calculations performed in reduced configuration spaces allow us to explore total energy curves as functions of proton-neutron pairing collective coordinates in selected $sd$-shell nuclei. In all cases studied, the self-consistent minima correspond to vanishing proton-neutron pairing, with energy increasing rapidly as proton-neutron pairing correlations are introduced. These results provide insight into the behavior of Gogny functionals under generalized HFB conditions and offer useful guidance for future developments.
We investigate proton-neutron ($pn$) pairing correlations within a generalized HFB framework using the Gogny D1S energy density functional. When $pn$ mixing is allowed, large single-particle bases trigger numerical instabilities in Gogny D1S due to its zero-range density-dependent term contribution to the $pn$ pairing field, whereas stable solutions are obtained with the Hamiltonian-based Brink-Boecker B1 interaction. Constrained HFB calculations in reduced spaces show that self-consistent minima with Gogny D1S systematically correspond to vanishing $pn$ pairing, with total energy rising rapidly as $pn$ pairing correlations are introduced. Furthermore, particle-number restored calculations significantly flatten these HFB energy curves, inducing $pn$ mixing across the collective coordinates within the variation after particle-number projection framework. These results provide key insights into the limitations and behavior of Gogny functionals under generalized HFB conditions, offering guidance for future functional re-parametrizations and beyond-mean-field studies.
\end{abstract}
\maketitle
%%%%%%%%%%%%
\section{Introduction}
%%%%%%%%%%%%%

The energy density functional (EDF) approach \cite{Bender03a,Robledo18a} is one of the most widely used theoretical methods to investigate the low-energy properties of atomic nuclei. The nuclear interaction is represented as a phenomenological energy functional that depends solely on one-body nucleon densities. Different families of functionals have been developed over the years such as the well-known Skyrme, Gogny, Fayans or relativistic functionals \cite{Bender03a,Decharge,Niksic11a,Reinhard2024a,Robledo18a}. 
The total energy of the system is then minimized by solving self-consistently the Hartree-Fock-Bogoliubov equations \cite{RS80a,Bender03a}, possibly under a set of constraints. In a pure mean-field (MF) picture, the Bogoliubov (or HFB) vacuum of lowest energy thus determined is taken as an approximation to the ground state of the nucleus. The method can also be extended to construct more correlated many-body wave functions by employing beyond-mean-field (BMF) techniques such as the quasi-particle random phase approximation \cite{Peru2014a}, the restoration of broken symmetries \cite{Sheikh2021a,Bally2021a} and the mixing of different HFB vacua within the framework of the generator coordinate method (GCM) \cite{Hill1953a,Griffin1957a,Egido16a,Bally2022b}. Ultimately, the quality of both the MF and BMF descriptions depends on the size of the variational space explored, which, in turn, is intimately related to the amount of symmetries that are allowed to be broken by the trial wave functions during the minimization procedure. For instance, the use of Bogoliubov vacua, which intrinsically break global gauge invariance associated with conserving a good number of particles, permits to account for pairing correlations.
Additionally, quadrupole (or higher multipoles) correlations can be included within a deformed HFB scheme that breaks the rotational invariance.

The parameters of EDFs are usually adjusted to reproduce, at a given level of approximation, selected experimental data (typically masses and charge radii) as well as pseudo-data such as properties of nuclear matter, specific single-particle gaps, fission barriers, and/or pairing properties, among other observables~\cite{Lalazissis1997,Chabanat1997,Garrido1999,Kortelainen2014,gognyoriginalpairing,Chappert2007,Chappert2015,Pillet2017a}. 
Consequently, the use of a specific parameterization beyond the approximation within which it was optimized may lead to inconsistencies. This occurs when degrees of freedom of the system not included in the wave functions used for the fit and validation of the parameters are explored.
Notably, most BMF calculations are performed employing parameterizations fitted at the MF level. This discrepancy can lead to minor issues, such as over binding or overestimation of nuclear deformation~\cite{Bender06a,Rodriguez15a}, or more severe ones, such as pathological behaviors of the functionals~\cite{Anguiano01a,Duguet09a,Bender09a,Lacroix09a}. From a pragmatic perspective, however, BMF calculations have proven an invaluable tool in the description of nuclear observables~\cite{Bender03a,Niksic11a,Robledo18a,Peru2014a,Egido16a}.

In this work, we investigate the behavior of energy functionals fitted considering Bogoliubov vacua with specific symmetry restrictions when employed in HFB calculations authorizing the breaking of additional symmetries. More specifically, we analyze the results from calculations with Gogny-type functionals (in particular, Gogny D1S~\cite{Berger1984} and a Hamiltonian version, Brink-Boecker B1 supplemented by a zero-range spin-orbit term~\cite{Blaizot1976,BRINK19671}) in a setting that allows the mixing of proton and neutron single-particle states in the HFB transformation. This generalization also enables the inclusion, at the mean-field level, of proton-neutron pairing correlations that have not been explored within this class of functionals. To perform this study, the numerical code \texttt{TAURUS} \cite{Sanchez_Fernandez2021,Bally2021b,Bally2024}, which was originally designed to handle general two-body Hamiltonians, has been extended to include zero-range density-dependent terms.

The article is organized as follows. Section~\ref{sec:theo} provides a brief overview of the theoretical framework and discusses the implications of proton-neutron mixing in the definition of HFB transformation. The one-body operators used to modulate the axial quadrupole deformation and the pairing of the HFB wave function are also introduced. Next, we analyze the stability of the Gogny D1S energy density functional when this new degree of freedom is included, and we compare the results with those obtained using the Brink-Boecker B1 Hamiltonian (Sec.~\ref{Stability}). Mean-field and particle-number projected energies and pairing-energy curves as functions of the aforementioned operators are then analyzed for selected nuclei in the $sd$ shell (Sec.~\ref{TES_delta}). Section~\ref{Summary} summarizes the main conclusions and outlines future perspectives. Finally, two appendices are included: one providing the explicit expressions for the fields arising from the density-dependent term when proton-neutron mixing is incorporated (App.~\ref{Fields}), and another (App.~\ref{Benchmark}) presenting a benchmark of the extended \texttt{TAURUS} code against widely used Gogny codes.

%%%%%%%%%%%%%%%%
\section{Theoretical framework}
\label{sec:theo}
%%%%%%%%%%%%%%%%

%%%%%%%%%%%%%%%%%%%%%%%%%%%%
\subsection{Basic principles}
%%%%%%%%%%%%%%%%%%%%%%%%%%%%
The HFB method has been extensively discussed in the nuclear structure literature \cite{RS80a,BR86a,Bender03a}. Therefore, in this section, we only recall the aspects most relevant for the present study. 

A Bogoliubov vacuum, $|\Phi(q)\rangle$, is defined as a product state of the form
\begin{equation}
  |\Phi(q)\rangle = \prod_{b} \beta_{b}(q) | 0 \rangle,
\end{equation}
where $| 0 \rangle$ is the bare vacuum and $\lbrace\beta_{b}(q);\beta^{\dagger}_{b}(q)\rbrace$ are quasi-particle annihilation and creation operators associated with the linear transformation
\begin{align}
\beta_{b_{\tau}}(q) &= \sum_{a_{\tau'}}U_{a_{\tau'}b_{\tau}}^*(q) c_{a_{\tau'}} + V_{a_{\tau'}b_{\tau}}^*(q) c^\dagger_{a_{\tau'}}, \\
\beta^{\dagger}_{b_{\tau}}(q) &= \sum_{a_{\tau'}}U_{a_{\tau'}b_{\tau}}(q)c^{\dagger}_{a_{\tau'}}+V_{a_{\tau'}b_{\tau}}(q)c_{a_{\tau'}},
\label{HFB_trans}
\end{align}
where $b_{\tau}$ denotes the index of the quasi-particle and the isospin $\tau=p/n$ is made explicit. 
The creation and annihilation operators $\lbrace c_{a} ; c^{\dagger}_{a}\rbrace$ define the working basis of the one-body Hilbert space, with $a$ being the index of the single-particle states.

The elements of the matrices $U(q)$ and $V(q)$ are the variational parameters of the problem that are determined by minimizing the total energy of the system under a set of constraints, which we collect under the label $q$. Typically, the constraints are applied on quantities of physical relevance such as the deformation or pairing content.

The HFB energy of state $|\Phi(q)\rangle$ can be expressed in terms of the one-body density $\rho(q)$, the one-body pairing tensor $\kappa(q)$, the Hartree-Fock field $\Gamma(q)$ and the pairing field $\Delta(q)$:
%\footnote{For simplicity of notation, the label $q$ will be omitted in the following expressions and reinstated when appropriate
\begin{align}
\rho_{a_{\tau}b_{\tau'}} (q) &=\langle\Phi(q)|c^{\dagger}_{b_{\tau'}}c_{a_{\tau}}|\Phi(q)\rangle \label{density_matrix},\\
\kappa_{a_{\tau}b_{\tau'}} (q) &=\langle\Phi(q)|c_{b_{\tau'}}c_{a_{\tau}}|\Phi(q)\rangle \label{pairing_tensor},\\
\Gamma_{a_{\tau}c_{\tau'}} (q) &=\sum_{b_{\tau''}d_{\tau'''}}\bar{v}_{a_{\tau}b_{\tau''}c_{\tau'}d_{\tau'''}}\rho_{d_{\tau'''}b_{\tau''}}(q) \label{HF_field},\\
\Delta_{a_{\tau}b_{\tau'}}(q) &=\frac{1}{2}\sum_{c_{\tau''}d_{\tau'''}}\bar{v}_{a_{\tau}b_{\tau'}c_{\tau''}d_{\tau'''}}\kappa_{c_{\tau''}d_{\tau'''}}(q) \label{Pair_field},\\
E^{\mathrm{HFB}} (q) &=\mathrm{Tr}\left[t\rho(q)\right]+\frac{1}{2}\mathrm{Tr}\left[\Gamma(q)\rho(q)\right] \\ 
                     &\phantom{=} -\frac{1}{2}\mathrm{Tr}\left[\Delta(q)\kappa^{*}(q)\right],
\end{align}\label{eq:HFBFieldsEnergyDefintions}
where $t_{ab}$ and $\bar{v}_{abcd}$ are the matrix elements of the kinetic-energy and two-body potential-energy operators, respectively.

% I comment this statement which is strange, and I think wrong: we could define a HFB transformation that does not mix single-particle states with different angular momentum.
%Equation~\eqref{HFB_trans} already shows that deformation and pairing correlations are introduced through the mixing of different single-particle angular momenta (quasi-particles do not have a well-defined angular momentum\bbfoot{I am not sure I understand this statement. What about spherical HFB?}) and through the mixing of creation and annihilation operators (the particle-number symmetry is not conserved), respectively. 

In most standard applications, the HFB transformation does not mix protons and neutrons, i.e., the matrix elements of $U(q)$ and $V(q)$ are non-vanishing only if $\tau = \tau'$. In that case, the one-body density matrix and pairing tensor exhibit a block diagonal structure with respect to the isospin. As a consequence, so do the Hartree-Fock and pairing fields, with matrix elements of the form $\Gamma_{a_{n}b_{p}}(q)$, $\Gamma_{a_{p}b_{n}} (q)$, $\Delta_{a_{p}b_{n}} (q)$, and $\Delta_{a_{n}b_{p}} (q)$ identically vanishing. This implies that certain proton-neutron components of the underlying interaction do not contribute to the energy. By contrast, when the HFB transformation allows proton-neutron mixing, there is no such isospin restrictions. As a result, the pairing energy
\begin{equation}
E^{\tau\tau'}_{\mathrm{pair}} (q) = \frac{1}{2}\sum_{a_{\tau}b_{\tau'}}\Delta_{a_{\tau}b_{\tau'}}(q) \kappa^{*}_{a_{\tau}b_{\tau'}}(q).
\label{pairing_ener}
\end{equation}
may contain contributions coming from proton-proton ($pp$), neutron-neutron ($nn$) and proton-neutron ($pn$) components.

%%%%%%%%%%%%%%%%%%%%%%%%%%%%
\subsection{TAURUS code and constraining operators}
%%%%%%%%%%%%%%%%%%%%%%%%%%%%
In this work, the \texttt{TAURUS} code is used to perform the calculations, as it allows for the use of general HFB transformations, with the restriction that the $U(q)$ and $V(q)$ matrices remain real. The working basis is the spherical harmonic oscillator, i.e., the operators $c^{\dagger}_{a}$ ($c_{a}$) create (annihilate) a nucleon in the state characterized by the quantum numbers $\lbrace a\rbrace \equiv \lbrace n_{a},l_{a},s_{a},j_{a},m_{j_{a}},\tau_{a}, m_{\tau_{a}}\rbrace$, which correspond to the principal oscillator quantum number, orbital angular momentum, spin, total angular momentum and its third component, isospin and its third component, respectively. 
The program is designed to define a genuine two-body Hamiltonian through its matrix elements $\bar{v}_{abcd}$ read as input. 
But as we wish to employ general density functionals, we have extended the code \texttt{TAURUS} to handle the rearrangement terms that arise when applying the variational principle to the density and pairing tensor. The most relevant expressions in this regard are provided in Appendix~\ref{Fields}.

Considering the constraints, the parameters $q$ explored in this work are the expectation values of operators that modulate the pairing content in the HFB vacuum. More precisely, we minimize the energy functional
\begin{align}
E'_{\mathrm{HFB}}\left[|\Phi(q)\rangle\right] = &\langle\Phi(q)|\hat{H}-\lambda_{N}\hat{N}-\lambda_{Z}\hat{Z}-\lambda_{q}\hat{Q}|\Phi(q)\rangle,
\label{HFB_eq}
\end{align}
where $\lambda_{N,Z,q}$ are Lagrange multipliers ensuring that $\langle \Phi(q)|\hat{N},\hat{Z},\hat{Q}|\Phi(q)\rangle = N,Z,q$, with $N(Z)$ being the neutron (proton) numbers. In this work, the operator $\hat{Q}$ refers to axial quadrupole deformation, using the usual $\beta_2$ parameter, and/or pairing-like operators. The latter are constructed from the pair creation (and annihilation) operators $[\hat{P}^{\dagger}]^{J_p T_p}_{M_{J_p}M_{T_p}}$ of two nucleons coupled to total angular momentum $J_{p}$, its third component $M_{J_{p}}$, total isospin $T_{p}$, and its third component $M_{T_{p}}$ as \cite{Bally2021b}
\begin{equation}
\begin{split}
\left[\hat{P}^{\dagger}\right]^{J_p T_p}_{M_{J_p}M_{T_p}} &= \sum_{\breve{a}} \left[\hat{P}^{\dagger}_{\breve{a}}\right]^{J_p T_p}_{M_{J_p}M_{T_p}} \\
&= \frac{1}{\sqrt{2}}\sum_{\breve{a}}\sqrt{2j_{a}+1}\left[c^{\dagger}_{\breve{a}}c^{\dagger}_{\breve{a}}\right]^{J_p T_p}_{M_{J_p}M_{T_p}}
\end{split}
\end{equation},
where we haved used the notation $\breve{a} \equiv  (n_a, l_a , j_a , s_a , \tau_a)$ and the creation operators are $J_p T_p$-coupled according to 
\begin{align}
\left[c^{\dagger}_{\breve{a}}c^{\dagger}_{\breve{b}}\right]&^{J_p T_p}_{M_{J_p}M_{T_p}} 
= \frac{\sqrt{1-\delta_{\breve{a} \breve{b}}(-1)^{J_p + T_p}}}{1+\delta_{\breve{a}\breve{b}}}\sum_{\substack{m_{j_{a}}m_{j_{b}} \\ m_{t_{a}}m_{t_{b}}}} c^{\dagger}_{a}c^{\dagger}_{b} \\
& \times \langle j_{a}m_{j_{a}}j_{b}m_{j_{b}}|J_p M_{J_p}\rangle \langle \tfrac{1}{2}m_{t_{a}}\tfrac{1}{2}m_{t_{b}}|T_p M_{T_p}\rangle . \nonumber 
\end{align}
The isovector channel ($J_p=0$, $T_p=1$) can be used to explore the usual $pp$-, $nn$-pairing correlations as well as a part of the $pn$ pairing. On the other hand, the isoscalar channel ($J_p=1$, $T_p=0$) is purely associated with $pn$ pairing.
More specifically, we define the parameters $\delta^{J_p T_p}_{M_{J_p}M_{T_p}}$ as
\begin{equation}
\delta^{J_p T_p}_{M_{J_p}M_{T_p}}= \frac12 \langle \Phi(q) | \left[\hat{P}\right]^{J_p T_p}_{M_{J_p}M_{T_p}} +
\left[\hat{P}^{\dagger}\right]^{J_p T_p}_{M_{J_p}M_{T_p}} | \Phi(q) \rangle.
\label{pair_op}
\end{equation} 
%%%%%%%%%%%%%%%%%%%%%%
\subsection{Effective interactions}
%%%%%%%%%%%%%%%%%%%%%
The most widely used Gogny interactions belong to the D1 family \cite{Berger1984}, whose expression is:
\begin{align}
\begin{split}
	   & \sum_{i=1,2} e^{-(\mathbf{r}_1-\mathbf{r}_2)^2/\mu_i^2} (W_i+B_iP^\sigma-H_iP^\tau-M_iP^\sigma P^\tau)\\
     & +\ t_3(1+x_0P^\sigma)\delta(\mathbf{r}_1-\mathbf{r}_2)\ \rho_{H}^{\xi}\left({\dfrac{\mathbf{r}_1+\mathbf{r}_2}{2}}\right) \\
	    &+iW_0(\mathbf{\sigma_1+\sigma_2}){\mathbf{k}}^\dagger \times \delta(\mathbf{r}_1-\mathbf{r}_2){\mathbf{k}},
	\label{eqn:D1Sinteraction}
\end{split}
\end{align}
%\begin{align}
%\begin{split}
%	   & \sum_{i=1,2} e^{-(\mathbf{r}_1-\mathbf{r}_2)^2/\mu_i^2} (W_i+B_iP^\sigma-H_iP^\tau-M_iP^\sigma P^\tau)\\
%	     &+\dfrac{e^2\ \delta(\tau^{1}_p,\  \tau^{
%	     2}_p)}{|\mathbf{r}_1-\mathbf{r}_2|}
%	     +iW_0(\mathbf{\sigma_1+\sigma_2}){\mathbf{k}}^\dagger \times \delta(\mathbf{r}_1-\mathbf{r}_2){\mathbf{k}}\\
%	     &+\ t_3(1+x_0P^\sigma)\delta(\mathbf{r}_1-\mathbf{r}_2)\ \rho_{H}^{\alpha}\left({\dfrac{\mathbf{r}_1+\mathbf{r}_2}{2}}\right),
%	\label{eqn:D1Sinteraction}
%\end{split}
%\end{align}
where $\mathbf{r}_1$, $\mathbf{r}_2$ are the position vectors of the nucleons, $P^{\sigma}$ and $P^{\tau}$ are the usual spin and isospin exchange operators, $\boldsymbol{\sigma}$ is the vector of Pauli matrices, $\mathbf{k}$ is the relative momentum, and $\rho_{H}(\mathbf{r})$ is the spatial density. At the MF level, this density is defined as the expectation value of the density operator in the HFB state. The remaining quantities $\{W_{i}, B_{i}, H_{i}, M_{i}, \mu_{i}, W_{0}, t_{3}, x_{0}, \xi\}$ are the interaction parameters adjusted following the protocols discussed in the Introduction~\cite{gognyoriginalpairing,Chappert2007,Chappert2015,Pillet2017a}. The first term is a finite-range central interaction consisting of the sum of two Gaussians, the second and the third terms correspond to zero-range density-dependent and spin-orbit contributions, respectively. In HFB calculations, the Coulomb interaction acting between protons is added.
The existence of two zero-range terms in the functional may lead to ultraviolet divergence issues in the pairing channel \cite{Bulgac02a}. In standard calculations without proton–neutron mixing, the spin–orbit pairing is expected to be small because it does not contribute to the spin-singlet/isospin-triplet channel. Moreover, the density-dependent term does not contribute to pairing in those cases either, since the parameter $x_{0}$ has been chosen equal to 1 for that purpose by D. Gogny who designed his functional only to treat pairing among like particles. The physical motivation under this choice of $x_0$ was the assumption that the proton-proton and neutron-neutron pairing interactions are only slightly renormalized by in-medium effects and are very close to the bare interaction. We remark, however, that recent ab initio calculations seem to indicate the importance of many-body effects \cite{Scalesi2026} in reproducing observables usually associated with pairing correlations. This is in agreement with previous findings obtained within the particle-vibration framework~\cite{Barranco2005}.
The antisymmetrized two-body matrix elements in the working basis, factorized into spatial, spin, and isospin parts, read
\begin{align}
\bar{v}^{DD}_{abcd}=t_{3}I^{DD}_{abcd}\left[S_{ac}S_{bd}\left(\delta_{\tau_{a}\tau_{c}}\delta_{\tau_{b}\tau_{d}}-x_{0}\delta_{\tau_{a}\tau_{d}}\delta_{\tau_{b}\tau_{c}}\right)+\right.& \\ \nonumber
\left.S_{ad}S_{bd}\left(x_{0}\delta_{\tau_{a}\tau_{c}}\delta_{\tau_{b}\tau_{d}}-\delta_{\tau_{a}\tau_{d}}\delta_{\tau_{b}\tau_{c}}\right)\right]&,
\end{align}
where $I^{DD}_{abcd}=\int\phi^{*}_{a}(\mathbf{r})\phi^{*}_{b}(\mathbf{r})\rho_{H}^{\alpha}(\mathbf{r})\phi_{c}(\mathbf{r})\phi_{d}(\mathbf{r})\,d\mathbf{r}$ is the integral of the density in coordinate space, which involves the single-particle wave functions $\left\{ \phi_{a} \right\}$ of the working basis, and $S_{ab}$ the spin matrix elements of the corresponding basis states. In the case where protons and neutrons are not mixed, the pairing tensor and pairing field satisfy $\tau_{a}=\tau_{b}\equiv\tau$ and $\tau_{c}=\tau_{d}\equiv\tau'$, so that the matrix elements entering the calculation of the pairing field simplifies as
\begin{align}
\bar{v}^{DD,pair}_{abcd}=t_{3}I^{DD}_{abcd}\left[S_{ac}S_{bd}(1-x_{0})+S_{ad}S_{bc}(x_{0}-1)\right]\delta_{\tau\tau'}&.
\end{align}
One sees that, as said before, the density-dependent term does not contribute to the $pp$ and $nn$ pairing channels. 
However, in the general case where the $pn$ mixing is authorized, the previous condition is clearly not satisfied. As a result, the zero range of the density-dependent term becomes a source of instability in the $pn$-pairing channel of the functional, and, on top of that, the intensity of this term is much larger than the spin-orbit one.
%\nath{In the following, we don't want to discuss the justification to introduce or not a density-dependence to the pn pairing channel. We only want to explore the consequence of the zero range form factor.}

Hence, to test the stability of the functionals with respect to the proton-neutron mixing channel, we also use the Brink-Boecker B1 interaction supplemented by a zero-range spin-orbit term~\cite{Blaizot1976}, in which the density-dependent term is absent in comparison with Gogny D1S and, in addition, $B_{i}=H_{i}=0$ is chosen in the central part. This pure Hamiltonian will allow us to check if the zero-range spin-orbit whose intensity is relatively small impacts the stability of the pairing channel. % \sout{the stability of the spin-orbit pairing channel}. 
%Finally, we also employ a density functional containing a density-dependent term similar to that of Gogny D1S (zero range), Yukawa-type central terms, and a finite-range spin-orbit term, namely the M3Y interaction (P2 parametrization).
%%%%%%%%%%%%%%%%
\section{Stability of the functionals}
\label{Stability}
%%%%%%%%%%%%%%%%

Before solving the HFB equation, Eq.~\eqref{HFB_eq}, the parameters of the working basis must be defined, namely the number of major oscillator shells, $N_{\mathrm{s.h.o.}}$, and the oscillator length, $b$. In the present work, a rather standard value of $b=1.005A^{-1/6}$~(fm) is adopted ($A$ is the mass number). In the limit where a sufficient number of oscillator shells is included, the HFB energies should become independent of these parameters, which is the usual criterion employed to assess the convergence of the solutions. We use the gradient method to perform the self-consistent calculations \cite{Bally2021a}.

To study how the inclusion of $pn$ mixing affects the solution of Eq.~\eqref{HFB_eq}, calculations are performed with an increasing number of oscillator shells and with two types of initial wave functions (seeds), with and without $pn$ mixing. It should be noted that, in our present numerical implementation, calculations with $N_{\mathrm{s.h.o.}}\geq8$ become computationally demanding. Consequently, we restrict this study 
to light nuclei that belong mainly to the $sd$ shell. Results with $N_{\mathrm{s.h.o.}}=10$ are used as limiting case to evaluate the convergence of the results.
%%%%%%%%%%%%%%%%%%%%
\begin{figure}[t]
    \centering
    \includegraphics[width=\columnwidth]{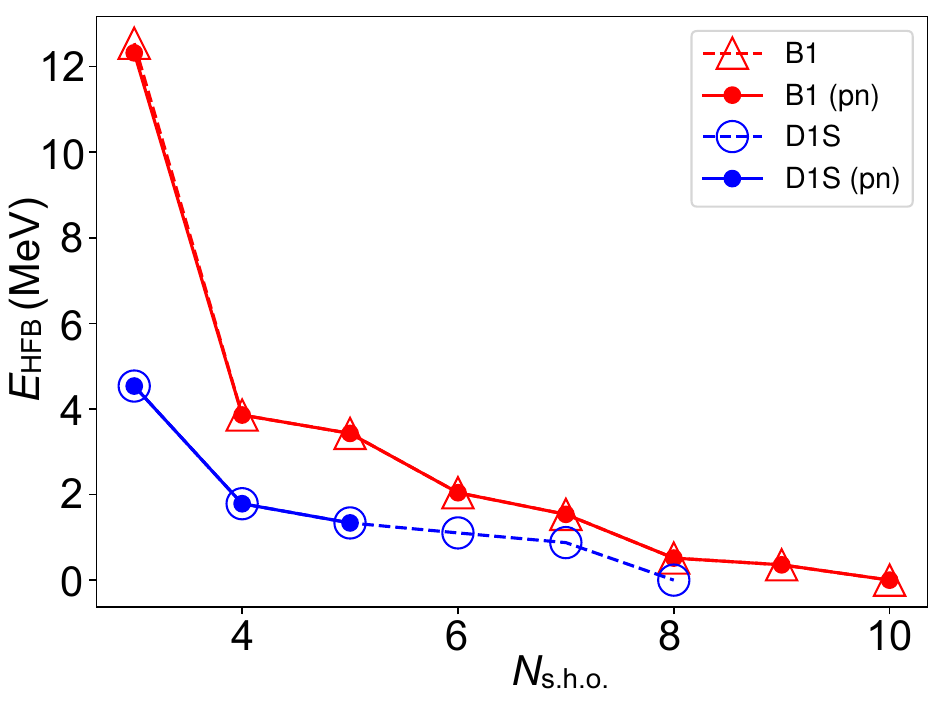}
    \caption{HFB energy as a function of the number of oscillator shells included in the working basis for the nucleus $^{20}$Mg, calculated with the Brink-Boecker B1 (red symbols) and Gogny D1S (blue symbols) interactions. Open symbols correspond to wave functions without $pn$ mixing, while filled symbols correspond to those allowing for $pn$ mixing. Energies are given relatively to the value obtained with the largest $N_{\mathrm{s.h.o.}}$ used in each case.}
    \label{Fig1}
\end{figure}
%%%%%%%%%%%%%%%%%%%%

Figure~\ref{Fig1} illustrates, as an example, the evolution of the HFB energy as a function of the number of oscillator shells for the nucleus $^{20}$Mg calculated with Brink-Boecker B1 and Gogny D1S parametrizations and with or without including $pn$ mixing in the seed wave function. The energies are given relatively to the final result obtained with the largest number of oscillator shells used for each interaction. The general trend is as expected: the energy decreases with the number of shells and eventually will saturate once convergence is achieved. For this nucleus, the B1 parametrization yields identical results regardless of whether the starting wave function allows $pn$ mixing. Since this functional lacks a density-dependent term but includes a zero-range spin–orbit term, it can be seen that the latter does not lead to any divergent behavior. A similar pattern is found for the Gogny D1S interaction when $pn$ mixing is not allowed. However, the most significant difference appears when $pn$ mixing is included in the minimization process. Up to $N_{\mathrm{s.h.o.}} = 5$, the results with and without mixing coincide; beyond this number of shells, the system becomes completely unstable and no converged solution is found.

%%%%%%%%%%%%%%%%%%%%
\begin{figure}[t]
    \centering
    \includegraphics[width=\columnwidth]{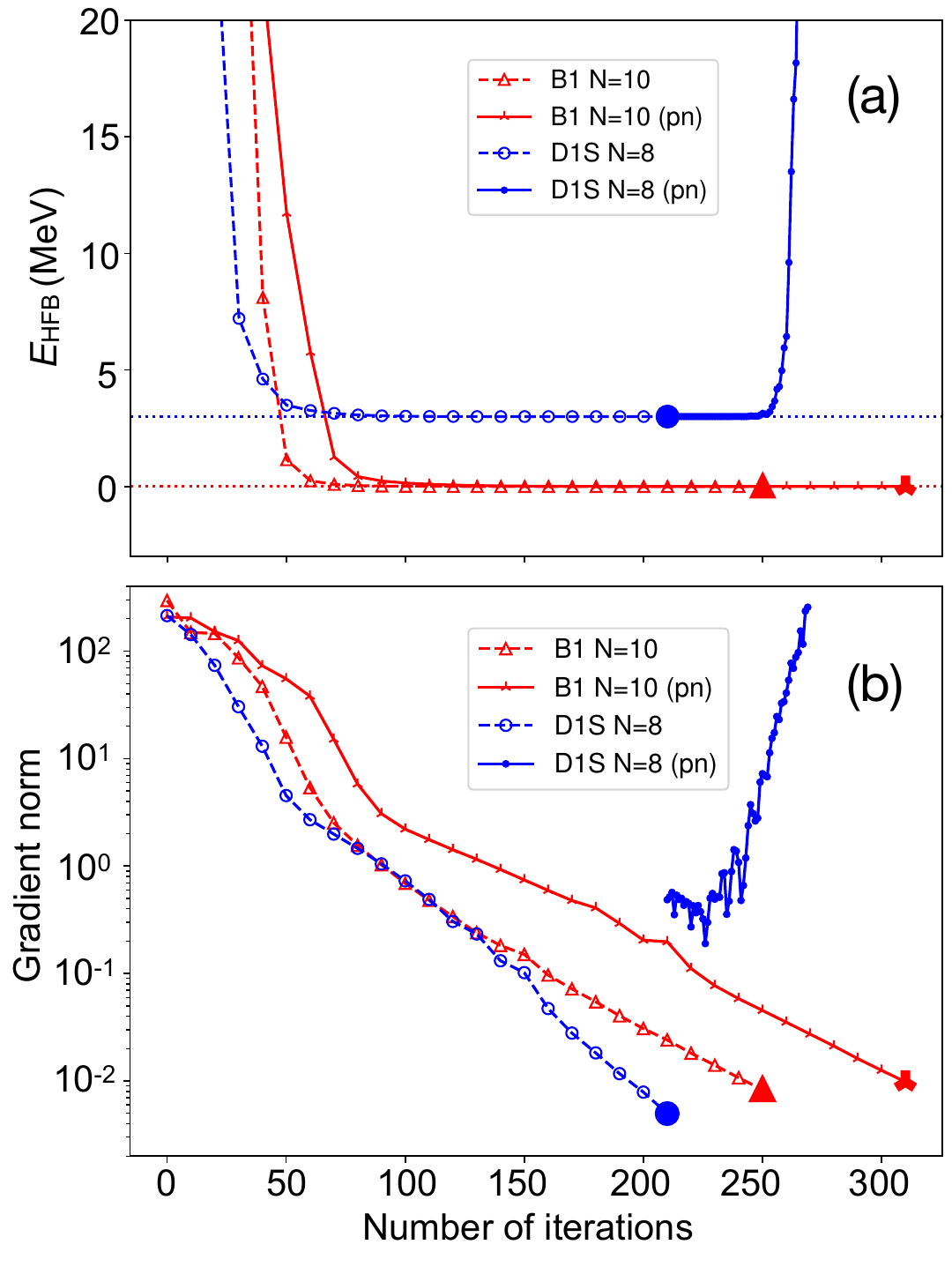}
    \caption{(a) HFB energy and (b) gradient as functions of the number of iterations of a particular computational run for the nucleus $^{20}$Mg, calculated with the Brink-Boecker B1 (red symbols) and Gogny D1S (blue symbols) interactions. Open symbols correspond to wave functions without $pn$ mixing, while filled symbols correspond to those allowing for $pn$ mixing. Energies are given relatively to the value obtained with the largest $N_{\mathrm{s.h.o.}}$ used in each case, and those corresponding to the Gogny D1S interaction are shifted by 3~MeV for clarity of presentation.}
    \label{Fig2}
\end{figure}
%%%%%%%%%%%%%%%%%%%%

This result is illustrated in Fig.~\ref{Fig2}, which shows the HFB energy and energy gradient as functions of the number of iterations in the minimization process for the B1 and D1S interactions, with $N_{\mathrm{s.h.o.}} = 10$ and 8, respectively. In calculations with the Brink-Boecker B1 interaction, both energy and gradient decrease steadily with the number of iterations, and a similar behavior is observed for Gogny D1S without $pn$ mixing. However, for Gogny D1S with $pn$ mixing, both quantities display erratic behavior that cannot be corrected by reducing the gradient step size or choosing a more suitable initial wave function. In this example, and to rule out a possible effect of an inappropriate choice of the initial state, Fig.~\ref{Fig2} shows a calculation that was first carried out without $pn$ mixing and a proper convergence was found after 215 iterations. Then, this converged wave function was slightly perturbed to include a small $pn$-mixing component. The continuation of the minimization process after this point with such a seed wave function showed oscillations in the gradient and a very large increase in the energy.  
Although this instability is exemplified here in the particular case of the nucleus $^{20}$Mg, it has been observed in our calculations for all nuclei within the $sd$-shell. Since the instability emerges when the configuration space is enlarged and the pairing channel associated with the (zero-range) density-dependent term becomes active, we attribute its origin to this component of the functional. In this regard, it is necessary to consider other functional forms, e.g., incorporating a finite-range density-dependent term as in the D2 Gogny interaction ~\cite{Chappert2007,Chappert2015} or fully finite range terms (including a tensor term) as in the  DG Gogny interaction ~\cite{Zietek2026,Zietek2023}.
%Since the $pn$ mixing affects both the Hartree-Fock field and the pairing field through the inclusion of proton-neutron interaction matrix elements, it is not possible to single out any specific term responsible for this instability, and it rather appears to be an issue related to the parameter-fitting protocol and/or the functional itself.
%%%%%%%%%%%%%%%%%%%%
\begin{figure}[tbh]
    \centering
    \includegraphics[width=\columnwidth]{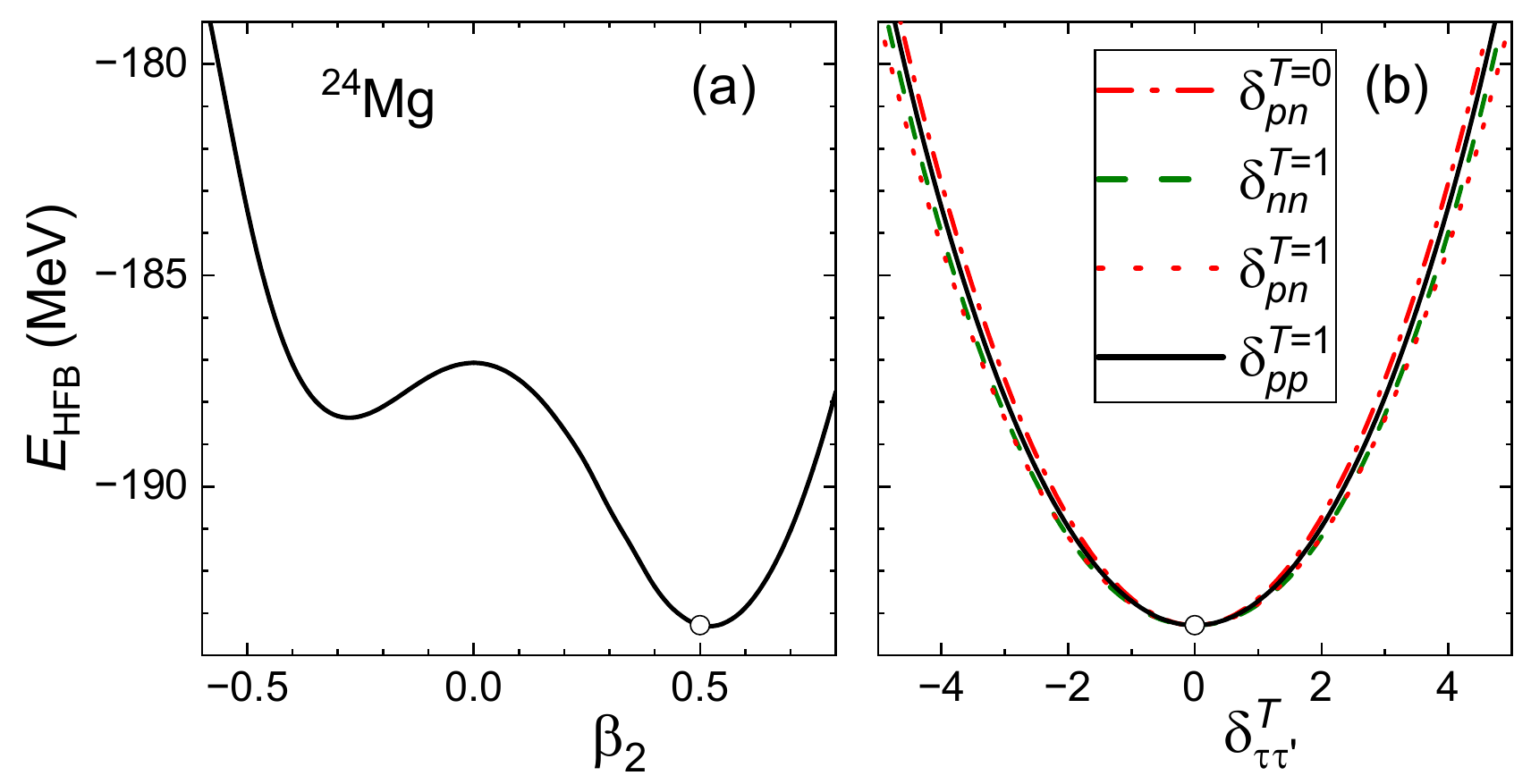}
    \caption{HFB energy as a function of (a) the axial deformation $\beta_{2}$ and, (b) pairing parameters $\delta^{T}_{\tau\tau'}$ calculated with Gogny D1S using 5 major oscillator shells. Continuous, dashed, dotted and dash-dotted lines corresponds to $\delta^{T=1}_{pp}$, $\delta^{T=1}_{nn}$, $\delta^{T=1}_{pn}$ and $\delta^{T=0}_{pn}$ parameters, respectively.}
    \label{Fig3}
\end{figure}
%%%%%%%%%%%%%%%%%%%%
\begin{figure*}[h!]
    \centering
%  \rotatebox{90}{%
    \includegraphics[width=0.8\textwidth]{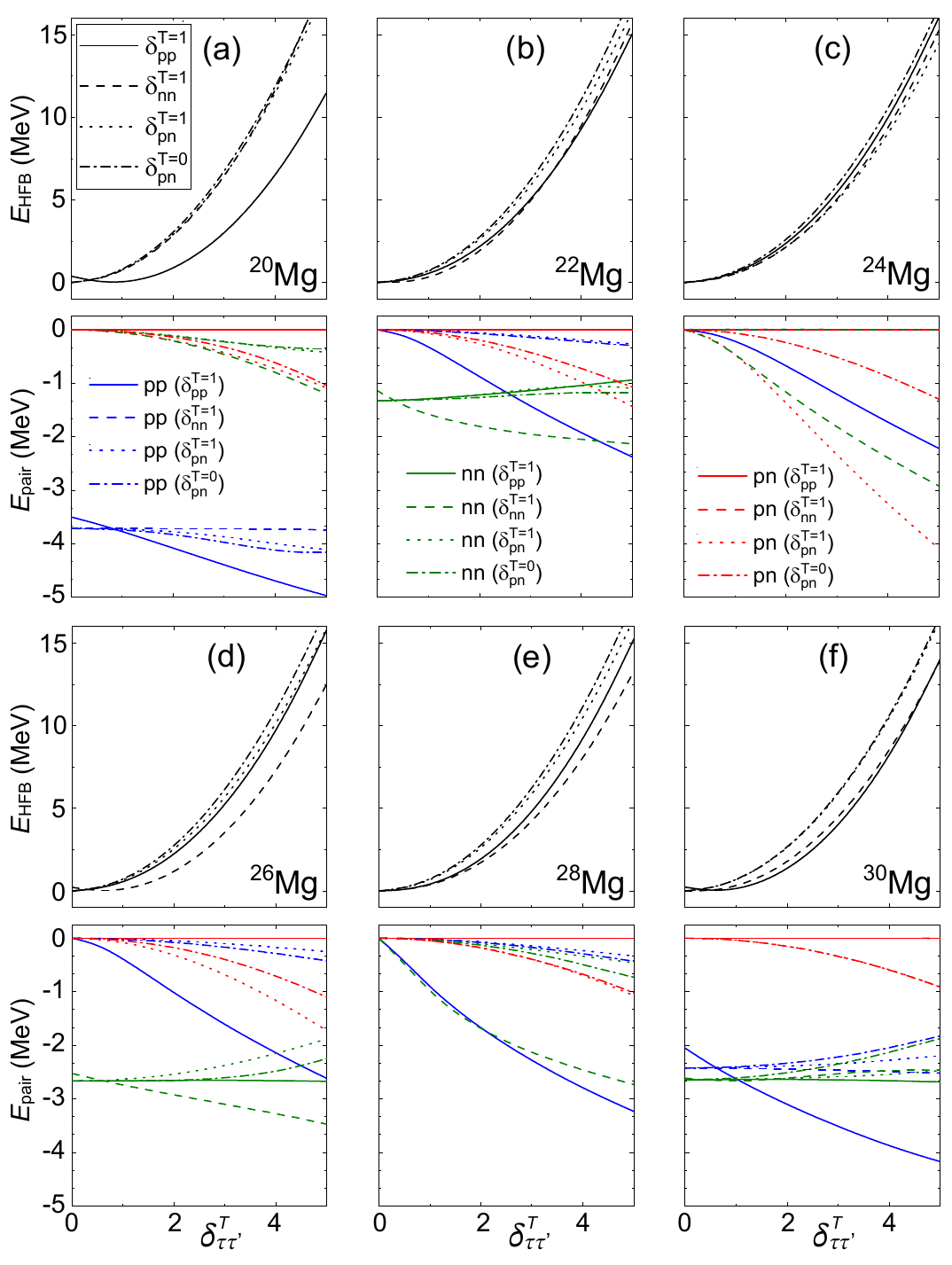}
%  }        
    \caption{HFB energy (top panel) and pairing energies (bottom panel) as a function of the pairing parameters $\delta^{T}_{\tau\tau'}$ for even-even Mg isotopes calculated with Gogny D1S using 5 major oscillator shells. The line style has the same meaning as in Fig.~\ref{Fig3}. Blue, green, and red colors are used for $E^{pp}_{\mathrm{pair}}$, $E^{nn}_{\mathrm{pair}}$, and $E^{pn}_{\mathrm{pair}}$, respectively.}
    \label{Fig4}
\end{figure*}
%%%%%%%%%%%%%%%%%%%%

%%%%%%%%%%%%%%%%%%%%
\begin{figure*}[t!]
    \centering
    \includegraphics[width=\textwidth]{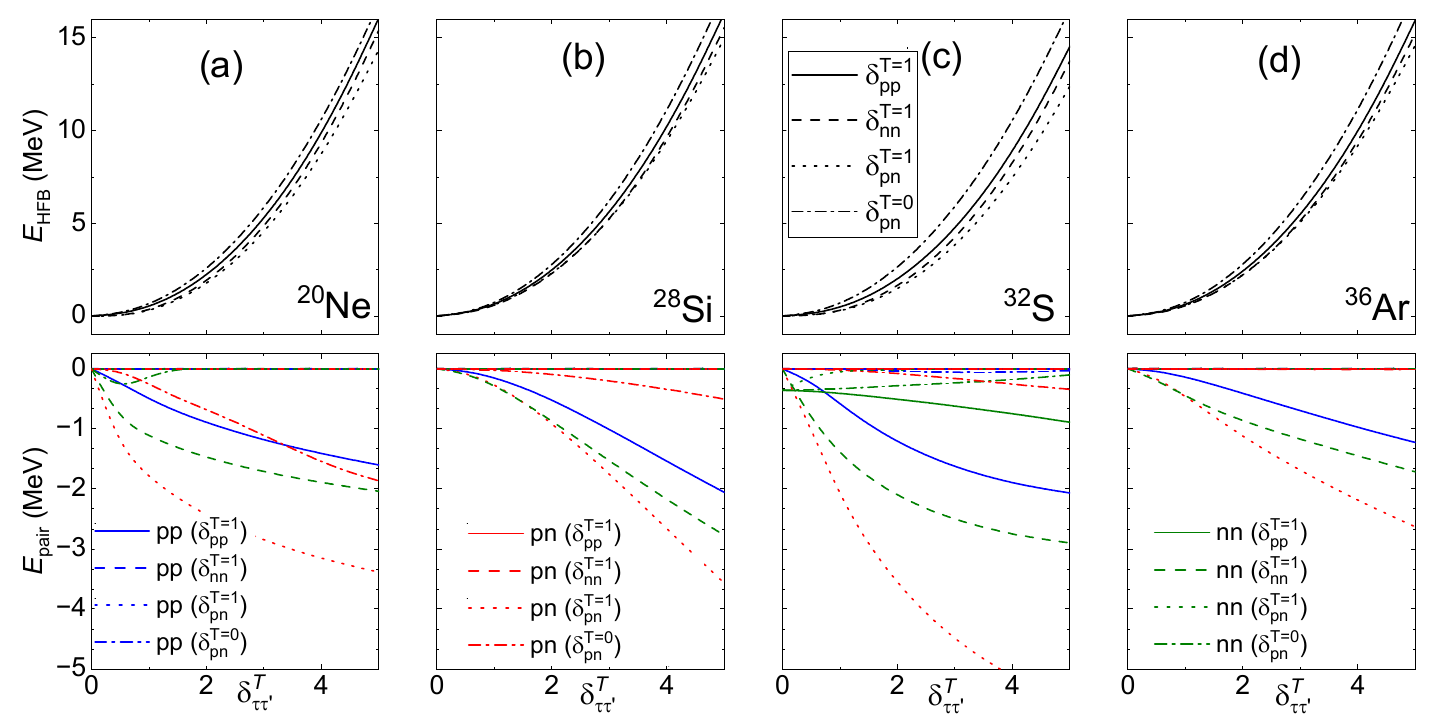}
    \caption{Same as Fig.~\ref{Fig4} but for $N=Z$ $sd$-shell nuclei.}
    \label{Fig5}
\end{figure*}
%%%%%%%%%%%%%%%%%%%%

%%%%%%%%%%%%%%%%%%%%
\section{Total energy curves along pairing degrees of freedom}
\label{TES_delta}
%%%%%%%%%%%%%%%%%%%%
Despite the fact that the Gogny D1S functional becomes unstable when the configuration space is large, for small working bases it is still interesting to study the MF energy as a function of the degrees of freedom $\delta^{T}_{\tau\tau'}$ and to analyze how the system responds to different types of pairing content. To this end, constrained calculations on these variables are performed for even-even nuclei in the $sd$ shell, using a working basis with five major oscillator shells. The use of a harmonic oscillator basis with a rather small number of shells can be viewed as an effective way to regularize the zero-range density-dependent term in the pairing channel (acting as an effective energy window for this channel). However, it should be emphasized that both the total and pairing energies depend on the basis size for the small number of major harmonic oscillator shells chosen. Therefore, their behavior with respect to the different pairing degrees of freedom depicts qualitative trends rather than strict quantitative predictions. The details of these calculations are as follows. First, the absolute HFB minimum is determined (assuming axial symmetry). The expectation values of the deformation and pairing operators for this wave function define the self-consistent values of these degrees of freedom. The HFB total energy curves (TECs) are obtained by keeping these values fixed, except for the $\delta^{T}_{\tau\tau'}$ degree of freedom under study, using constrained HFB calculations.
Figure~\ref{Fig3} shows an example of this type of calculation. In Fig.~\ref{Fig3}(a), it is observed that the absolute minimum for the nucleus $^{24}$Mg corresponds to a prolate deformed configuration with $\beta_{2}=0.5$ (white bullet). In addition, this wave function has all $\delta^{T}_{\tau\tau'}=0$. These expectation values define the self-consistent values. In Fig.~\ref{Fig3}(b), the HFB energy is shown when constraining both $\beta_{2}$ and all $\delta^{T}_{\tau\tau'}$ to their self-consistent values, except for the one that is varied and displayed on the $x$-axis. Depending on the type of $\delta^{T}_{\tau\tau'}$ being explored, the energies are represented by solid ($\delta^{T=1}_{pp}\equiv\delta_{pp}$), dashed ($\delta^{T=1}_{nn}\equiv\delta_{nn}$), dotted ($\delta^{T=1}_{pn}$), and dash-dotted ($\delta^{T=0}_{pn}$) lines. Obviously, the absolute minimum of this TEC corresponds to the (axial) self-consistent solution obtained without constraints. Moreover, it is observed that $\delta^{T}_{\tau\tau'}$ can take both positive and negative values, since this variable is not restricted to positive values as would be the case where a genuine two-body operator is used to count the number of nucleon pairs coupled to a given $JT$.
The result obtained for $^{24}$Mg (and for all other nuclei studied in this work) is that the energy is symmetric around the minimum for any type of $\delta^{T}_{\tau\tau'}$, in this case around $\delta^{T}_{\tau\tau'}=0$ in all channels. In the following figures, only values with $\delta^{T}_{\tau\tau'} \geq 0$ are shown. Furthermore, the energy increases rapidly with increasing $\delta^{T}_{\tau\tau'}$, and no local minimum is observed. The energy increase is steeper in the $T=0$ channel, followed by the $pp$ channel, then the $nn$ channel, while the softer curve corresponds to the isovector $pn$ pairing channel.

To analyze the evolution of these curves as a function of the proton-neutron asymmetry of the system, Figs.~\ref{Fig4} and~\ref{Fig5} show these TECs (upper panels) and the corresponding $nn$, $pp$, and $pn$ pairing energies defined in Eq.\ref{pairing_ener} (lower panels) for the magnesium isotopic chain and for $N=Z$ nuclei, respectively. In addition to the style of the line that identifies the type of $\delta^{T}_{\tau\tau'}$ being explored mentioned above, pairing energies [Eq.~\eqref{pairing_ener}] are distinguished by color: blue, green, and red for $pp$, $nn$, and $pn$ pairing, respectively.
As a general rule, the self-consistent minima correspond to $\delta^{T=0,1}_{pn}=0$ and vanishing $pn$ pairing energy. Moreover, variations in $\delta^{T}_{\tau\tau'}$ mainly lead to an increase in the pairing energy of the channel being explored. For example, varying $\delta_{pp}$ ($\delta_{nn}$) produces a significant increase in the $pp$ ($nn$) pairing energy. On the other hand, when exploring $\delta_{pp}$ or $\delta_{nn}$, the pairing energies in the opposite channels ($nn$ and $pp$, respectively) remain practically constant, as do the $pn$ pairing energies, which are zero over the entire range of $\delta_{pp/nn}$. This behavior changes when constraining $\delta^{T=0,1}_{pn}$, where, in addition to an increase in the $pn$ pairing energy, the $pp$ and $nn$ pairing energies also change, either enhancing or reducing pairing correlations in those channels.

Focusing on the TECs of the magnesium isotopes (upper panel of Fig.~\ref{Fig4}), one observes that, except for the $N=Z$ nucleus and the neutron shell closure at $N=8$, the curves associated with $nn$ and $pp$ pairing are broader than those of the $pn$ channels, and their corresponding pairing energies are larger than the $pn$ ones. In addition, the isoscalar $pn$ channel exhibits steeper curves than the isovector one. Besides the repulsive nature of the density-dependent term in this channel, this behavior could also be related to the absence of a tensor term in the Gogny D1S functional. Although the tensor force is strongly attractive in the bare $S=1,T=0$ channel, its effect in an EDF framework is modified by in-medium renormalization. Nevertheless, omitting tensor forces leaves out non-central couplings that might favor $T=0$ pairing. A detailed study of how effective tensor interactions affect proton-neutron pairing can be addressed in future work using the recently developed DG Gogny functional, which includes finite-range tensor terms~\cite{Zietek2023,Zietek2026}. This behavior changes slightly for $N=Z$ nuclei in the $sd$ shell (see Fig.~\ref{Fig5} and Fig.~\ref{Fig4}(c)), where the lowest-energy TECs correspond to those exploring $\delta^{T=1}_{pn}$, followed by $\delta_{nn}$, $\delta_{pp}$, and finally $\delta^{T=0}_{pn}$. This can be understood by examining the pairing energies in the different channels, where the $pn$ pairing energy $E^{pn}_{\mathrm{pair}}$ is largest when exploring $\delta^{T=1}_{pn}$, even larger than $E^{nn}_{\mathrm{pair}}$ when exploring $\delta_{nn}$. In the case of $\delta_{pp}$, the Coulomb term introduces an anti-pairing effect, leading to $E^{pp}_{\mathrm{pair}} < E^{nn}_{\mathrm{pair}}$ even for $N=Z$ systems. In any case, the energy increase is not sufficiently different among the various channels to justify neglecting any of them \textit{a priori} in BMF calculations that include fluctuations in the pairing degrees of freedom (e.g., within the generator coordinate method). 

The \texttt{TAURUS} suite enables the inclusion of beyond-mean-field effects via symmetry restoration (particle number, angular momentum, and parity) and configuration mixing within the generator coordinate method, employing generic Hamiltonians with up to two-body interactions. However, the implementation of density functionals within beyond-mean-field methods in the \texttt{TAURUS} suite is still currently a work in progress. Nevertheless, to investigate the changes induced in the TECs as a function of the pairing parameters when beyond-mean-field effects are included, we use the Brink-Boeker B1 interaction introduced in Sec.~\ref{sec:theo}. Once again, single-particle orbits up to the $gds$ shell are considered. Since the $\delta^{T}_{\tau\tau'}$ parameters primarily break particle-number symmetry, we examine how the geometry of the TECs changes upon projection onto a good particle number. We also study the difference between projection after variation (PN-PAV) and the minimization of the particle-number-projected energy (PN-VAP), while preserving axial symmetry.
%%%%%%%%%%%%%%%%%%%%
\begin{figure}[t!]
    \centering
    \includegraphics[width=\columnwidth]{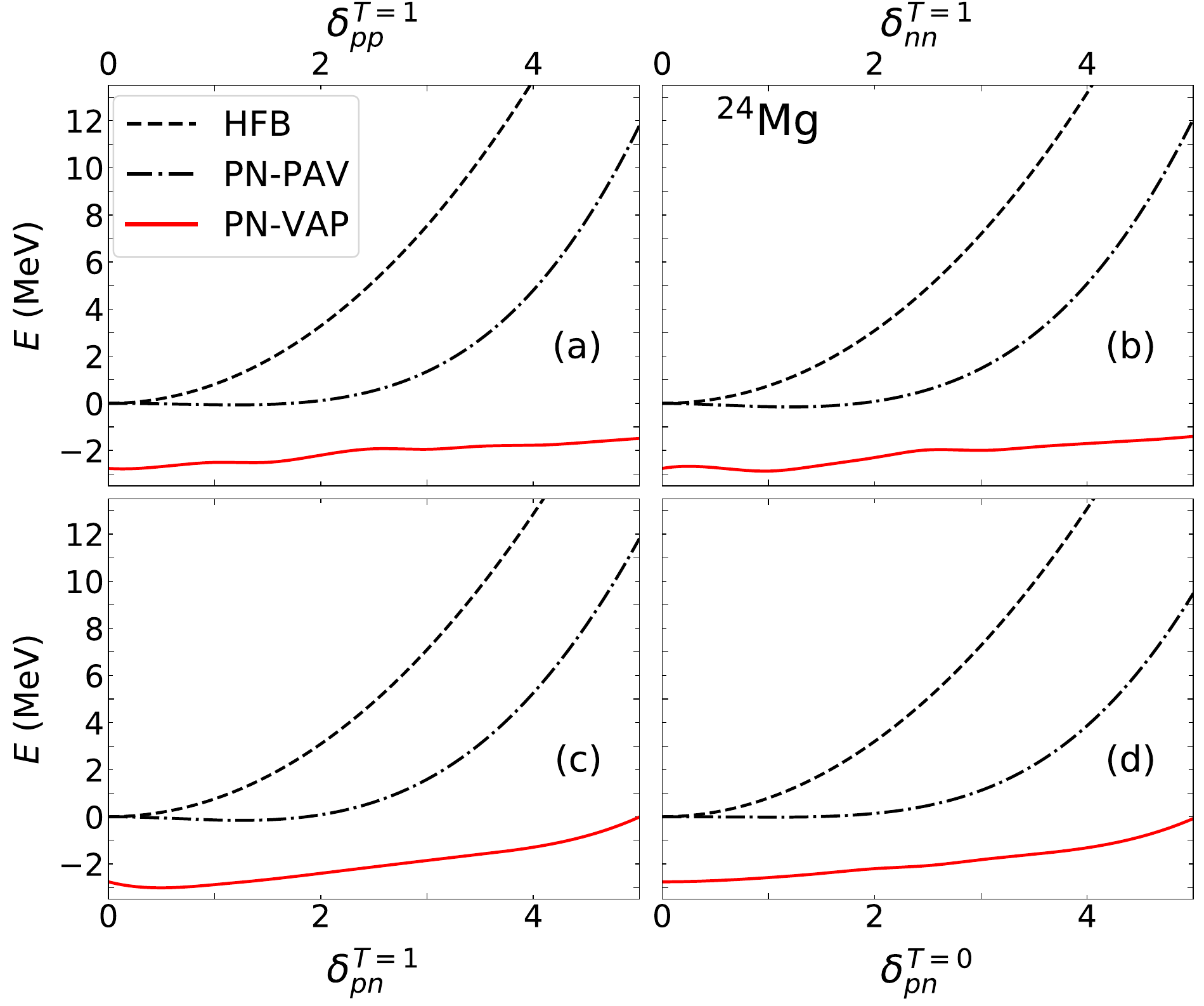}
    \caption{HFB, PN-PAV (dotted and dashed-dotted black lines, respectively) and PN-VAP (continuous red line) energies as a function of the pairing parameters $\delta^{T}_{\tau\tau'}$ computed with B1 interaction ($N_{\mathrm{s.h.o}}=5$) for $^{24}$Mg.}
    \label{Fig6}
\end{figure}
%%%%%%%%%%%%%%%%%%%%
In Fig.~\ref{Fig6} we display the HFB, the corresponding PN-PAV, and the PN-VAP TECs for the $^{24}\mathrm{Mg}$ nucleus ($N=Z$) evaluated across the different pairing parameters, $\delta^{T}_{\tau\tau'}$. Similar results are obtained in other nuclei in the $sd$-shell region (not shown). The deformation has been fixed at the value corresponding to the HFB energy minimum, which in this case is $\beta_{2}=0.5$. The mean-field result exhibits a behavior very similar to that found with the Gogny D1S interaction; that is, the energy increases significantly away from the minimum, showing a comparable curvature across all $\delta^{T}_{\tau\tau'}$ channels. A similar trend is observed for the PN-PAV results, although correlation energy is gained due to symmetry restoration, except at $\delta^{T}_{\tau\tau'}=0$, which corresponds to a HF state in all cases. Nonetheless, the behavior of these curves is significantly flatter and the minima are found at values with $\delta^{T}_{\tau\tau'}\neq 0$. Finally, the PN-VAP yields the lowest energy values and an even flatter behavior. Furthermore, unless the seed wave function is restricted to exclude proton-neutron mixing, the PN-VAP projection induces particle-number symmetry breaking and proton-neutron mixing in the HFB transformation across the entire range of $\delta^{T}_{\tau\tau'}$, even at $\delta^{T}_{\tau\tau'}=0$. This highlights the relevance of the PN-VAP method to dynamically capture not only $pp$ and $nn$ pairing correlations, but also $pn$ ones, even when they are absent at the unprojected mean-field level. Furthermore, the fact that the particle-number projected energy curves are remarkably flat along these collective coordinates suggests that pairing fluctuations around the minimum could yield substantial correlation energies. This underscores their crucial role in future beyond-mean-field calculations incorporating configuration mixing within the projected generator coordinate method (PGCM).
%As a final remark,  similar conclusions can be drawn from calculations performed with Brink-Boecker B1 interaction (not shown). 
%%%%%%%%%%%%%%%%
\section{Summary}
\label{Summary}
%%%%%%%%%%%%%%%%
In this work, we have explored the inclusion of proton-neutron pairing correlations at the MF level within the Hartree-Fock-Bogoliubov framework using Gogny-type functionals. By allowing for proton-neutron mixing in the quasi-particle transformation, both isovector ($T=1$) and isoscalar ($T=0$) pairing channels can be treated on an equal footing with like-particle pairing. To perform such calculations, the \texttt{TAURUS} code has been extended to incorporate density-dependent Gogny energy density functionals in the presence of proton-neutron mixing.  As a first step, the reliability of the extended code has been assessed through a benchmark against standard HFB calculations without proton-neutron mixing. In the absence of proton-neutron mixing, full agreement with previously validated implementations is found, providing confidence in the numerical correctness of the generalized framework (see App.\ref{Benchmark}).
A detailed analysis of the numerical stability of the Gogny D1S functional reveals that the inclusion of proton-neutron mixing leads to instabilities when sufficiently large harmonic-oscillator bases are employed, as previously discussed and expected. These instabilities manifest themselves through a non-convergent behavior of the HFB energy and gradient during the minimization procedure. In contrast, calculations performed with the Hamiltonian-based Brink-Boecker B1 interaction, which does not contain density-dependent terms, remain stable under the same generalized HFB conditions. This comparison strongly suggests that the density-dependent term of the Gogny D1S functional plays a central role in the observed instabilities once the additional $pn$-pairing channel is activated. Future studies concerning these instabilities should be performed in finite-range density dependent Gogny interactions such as D2~\cite{Chappert2015} and DG~\cite{Zietek2026} functionals.
Although the instability of Gogny D1S prevents fully converged calculations in large configuration spaces, meaningful information can still be extracted from constrained HFB calculations performed in reduced bases. While the absolute total and pairing energies remain basis-dependent in such truncated spaces, these calculations allow us to extract qualitative trends. Total energy curves as functions of pairing collective coordinates were computed for selected even-even nuclei in the $sd$ shell. In all cases studied, the self-consistent HFB minima correspond to vanishing proton-neutron pairing correlations. Moreover, fluctuations in the proton-neutron pairing degrees of freedom, both in the isovector and isoscalar channels, lead to a rapid and symmetric increase of the total energy around the minimum. The isoscalar pairing channel is found to be systematically stiffer than the isovector one, while like-particle pairing channels generally produce broader energy curves, except in $N=Z$ nuclei, where isovector proton-neutron pairing becomes comparatively more competitive. In addition, the observed stiffness of the isoscalar channel highlights the potential role of tensor interactions, which are absent in standard Gogny D1 parametrizations.
The present results demonstrate that widely used Gogny functionals, optimized within restricted HFB frameworks without proton-neutron mixing, may not be suitable for generalized MF calculations that explicitly include proton-neutron pairing correlations. Additionally, the symmetry-restored calculations emphasize the key role of the PN-VAP approach, which naturally induces proton-neutron mixing and leads to significantly flatter total energy landscapes compared to the mean-field case. This dynamic inclusion of $pn$ pairing correlations highlights the necessity of incorporating fluctuations in both isovector and isoscalar pairing degrees of freedom in future configuration-mixing studies. Future work will focus on the construction and optimization of functionals tailored to generalized HFB calculations, as well as on the inclusion of proton-neutron pairing fluctuations within symmetry-restored and generator coordinate method approaches. 
\section*{Acknowledgments}
We acknowledge funding from the Spanish MICIN under PID2021-127890NB-I00 and PID2024-159559NB-C22. This work has received funding from the European Research Council under the European Union’s Horizon Europe Research and Innovation Programme (Grant Agreement No.\ 101162059).
\appendix

\section{proton-neutron density-dependent fields.}\label{Fields}
As mentioned, the density-dependent interactions cannot be completely expressed as matrix elements due to the Hamiltonian symmetry breaking from different spatial deformations (for example, nuclear octupolar deformation would have parity-breaking density-dependent matrix elements), not compatible with the interaction setup implemented in \texttt{TAURUS}. Also, it is not possible to obtain the rearrangement as matrix elements.

Therefore, the density-dependent term is implemented in the code as the final field expression, adding the different spatial densities before evaluating the matrix element integral.

In the following, we use the notation for the single-particle wave functions (where the Greek letter explicitly represents $m_j$, but if not specified, also the other quantum numbers represented by the Latin letter)
\begin{align}
    \begin{split}
\quad&\varphi_{\alpha}^\tau(\mathbf{r})=\langle{\mathbf{r}|\alpha}\rangle^{(\tau)}\equiv  \langle{\mathbf{r}|n_al_as_aj_a,m_{j_a}\equiv\alpha}\rangle^{(\tau)}\\
&=R_{n_al_a}(r)\sum_{m_{l_a}m_s} C_{l_am_{l_a}1/2 m_s}^{j\alpha} Y_{l_am_{l_a}}(\Omega)|m_s\rangle|\tau\rangle\\
    \end{split}
\end{align}

The radial part $R_{nl}(r)=\overline{R}_{nl}(r) e^{-\frac{r^2}{2b^2}}$, with the harmonic oscillator length  $b$:
\begin{equation}
\label{eqn:radial1B}
\overline{R}_{nl}(r) = \dfrac{1}{b^{3/2}} \sqrt{\frac{2^{n+l+2}\ n!}{\sqrt{\pi}(2n+2l+1)!!}}\ \left({\frac{r}{b}}\right)^l \mathcal{L}_{n}^{l+1/2}\left({(r/b)^2}\right)
\end{equation}
For the integral of zero-range, the functions group in pairs, the radial two-body part can be expressed directly by the product of function [ see Eq.~\eqref{eqn:radial1B}] or as a polynomial from such functions. Both are computationally equivalent. The angular parts are defined from the spherical harmonic products
\begin{align}
\begin{split}
\label{eqn:SphericalHarmonicsDUAL}
&\chi_{l_am_{l_a}l_bm_{l_b}}^0(\Omega)={Y}^*_{l_am_{l_a}}(\Omega)Y_{l_bm_{l_b}}(\Omega)\\
&\chi_{l_am_{l_a}l_bm_{l_b}}^1(\Omega)=Y_{l_am_{l_a}}(\Omega)Y_{l_bm_{l_b}}(\Omega)\\
&\chi_{l_am_{l_a}l_bm_{l_b}}^2(\Omega)={Y}^*_{l_am_{l_a}}(\Omega){Y}^*_{l_bm_{l_b}}(\Omega)\\
\end{split}
\end{align}
The products in the integral are combined in terms identified by the different spin components $m_s$ and the only orbital's third component given by $m_l\!=\!m_j-m_s$
\begin{align}
\begin{split}
\label{eqn:AngFuncDUALonMs}
&{}^0\Xi^{\ \alpha\gamma\ (m_s,m_s')}_{l_am_{l_a}l_cm_{l_c}}=C_{l_am_{l_a}1/2\ m_s}^{j_a\alpha}C_{l_cm_{l_c} 1/2\ m_s'}^{j_c\gamma}\ \chi_{l_am_{l_a}l_cm_{l_c}}^0\\
&{}^1\Xi^{\ \alpha\beta\ (m_s,m_s')}_{l_am_{l_a}l_bm_{l_b}}=C_{l_am_{l_a}1/2\ m_s}^{j_a\alpha}C_{l_bm_{l_b}1/2\  m_s'}^{j_b\beta}\ \chi_{l_am_{l_a}l_bm_{l_b}}^1\\
&{}^2\Xi^{\ \alpha\beta\ (m_s,m_s')}_{l_am_{l_a}l_bm_{l_b}}=C_{l_am_{l_a}1/2\ m_s}^{j_a\alpha}C_{l_bm_{l_b}1/2\  m_s'}^{j_b\beta}\ \chi_{l_am_{l_a}l_bm_{l_b}}^2\\
\end{split}
\end{align}
with  $C_{l_am_{l_a}1/2\ m_s}^{j_a\alpha}$ the Clebsch-Gordan coefficients.

In the following, for the local-density expression $\langle{\alpha|\delta(\mathbf{r})|\beta}\rangle=\varphi_i^*(\mathbf{r})\varphi_j(\mathbf{r})=R_{ab}(r)\mathcal{Y}_{\alpha\beta}({\Omega})\ \delta^\tau_{ij}$, specifying the angular function
\begin{align}
    \begin{split}
        \mathcal{Y}_{\alpha\beta}({\Omega})&=
\sum_{m_s} C_{l_am_{l_a}1/2\ m_s}^{j_a\alpha}
C_{l_bm_{l_b}1/2\ m_s}^{j_b\beta}\ {Y}^*_{l_am_{l_a}}Y_{l_bm_{l_b}}\\
&={}^0\Xi^{j_a\alpha\ j_b\beta\ (+,+)}_{l_a(\alpha-1/2)l_b(\beta-1/2)} + {}^0\Xi^{j_a\alpha\ j_b\beta\ (-,-)}_{l_a(\alpha+1/2)l_b(\beta+1/2)}\\
    \end{split}\label{eq:angular2BFunction_definition}
\end{align}
From these expressions, we also define bulk-density fields in terms of spin and isospin contributions: $A_{(m_s,m_s')}$ for the HF parts and two $B_{(m_s,m_s')}$ for the pairing fields.  
\begin{align}
   \begin{split}
\rho^{(\tau_a\tau_c)}(\mathbf{r})&=\sum_{\beta(\tau_c)\delta(\tau_a)}\ \rho_{\delta\beta}\ 
\mathcal{Y}_{\beta\delta}(\Omega)R_{bd}(r)
\\
A^{(\tau_a\tau_c)}_{(m_s,m_s')}(\mathbf{r}) &= \sum_{\beta(\tau_c)\delta(\tau_a)}\ \rho_{\delta\beta}\ {}^0\Xi^{\ \beta\delta\ (m_s',m_s)}_{l_bm_{l_b}l_dm_{l_d}}(\Omega) R_{bd}(r)
\\
{}^1B^{(\tau_a\tau_b)}_{(m_s,m_s')}(\mathbf{r}) &= \sum_{\gamma(\tau_a)\delta(\tau_b)}\ \kappa_{\gamma\delta}\ {}^1\Xi^{\ \gamma\delta\ (m_s,m_s')}_{l_cm_{l_c}l_dm_{l_d}}(\Omega) R_{cd}(r)
\\
{}^2B^{(\tau_1\tau_2)}_{(m_s,m_s')}(\mathbf{r}) &= \sum_{\alpha(\tau_1)\beta(\tau_2)}\ \kappa^*_{\alpha\beta}\ {}^2\Xi^{\ \alpha\beta\ (m_s,m_s')}_{l_am_{l_a}l_bm_{l_b}}(\Omega) R_{ab}(r)
\\
\rho^{0}(\mathbf{r})=\rho^{(pp)} + &\rho^{(nn)}\qquad A^{0}_{(m_s,m_s')}=A^{(pp)}_{(m_s,m_s')} + A^{(nn)}_{(m_s,m_s')},
   \end{split} 
\label{eqn:BulkFields4PNspace}
\end{align}
where $\rho_{\alpha\beta}$ and $\kappa_{\alpha\beta}$ are the density matrix and the pairing tensor as defined in Eqs.~\eqref{density_matrix}-~\eqref{pairing_tensor}.
The one-body spatial density is evaluated in second quantization with the density matrix $\rho_{\varepsilon\eta}$
\begin{align}
\label{eqn:deltaDiracME}
\begin{split}
    \hat{\rho}(\mathbf{r})&=\sum_{\varepsilon\eta} d_{\varepsilon\eta}(\mathbf{r})\ c_\varepsilon^\dagger c_\eta\\
    d_{\varepsilon\eta}(\mathbf{r})&=\langle{\varepsilon|\delta (\mathbf{r}-\hat{\mathbf{r}})|\eta}\rangle=\varphi_\varepsilon^*(\mathbf{r})\varphi_\eta(\mathbf{r})\ \delta^\tau_{\varepsilon\eta}
\end{split}
\end{align}
From the HF and pairing fields, Eq.~\eqref{eq:HFBFieldsEnergyDefintions}, and the expression of the matrix element $\overline{v}^{DD}_{\alpha\beta\gamma\delta}$ in the spherical basis, we can get the different fields, where the usual \textit{pp-nn} fields are denoted here as $\Gamma^{\tau}$ and $\Delta^{\tau}$, and the new \textit{pn} terms are $\Gamma^{\tau\tau'}$ and $\Delta^{\tau\tau'}$ ($\tau\neq\tau'$). There is the spin factor is $X_{abcd} = \delta_{ac}^{\tau}\delta_{bd}^{\tau'} - x_0\delta_{ad}^{\tau}\delta_{bc}^{\tau'}$
\begin{widetext}
\begin{equation}
    \overline{v}^{DD}_{\alpha\beta\gamma\delta} = t_3\int dr^3 \rho^\xi(\mathbf{r}) R_{abcd}(r)
\left({
\mathcal{Y}_{\alpha\gamma}\mathcal{Y}_{\beta\delta}
X_{abcd} - 
\mathcal{Y}_{\alpha\delta}\mathcal{Y}_{\beta\gamma} X_{abdc}
}\right)\left({\Omega}\right)
\end{equation}
\begin{align}
    \begin{split}
\Gamma_{\alpha\gamma}^{(\tau)}=\  t_3\int& d\mathbf{r}\ R_{ac}\ \rho^\xi(\mathbf{r})\ \times
\\ 
\mathcal{Y}_{\alpha\gamma}(\Omega)\ \left({\rho^0(\mathbf{r})
-\ x_0\  \rho^{(\tau)}(\mathbf{r})}\right)
-&\sum_{m_s\ m_s'} {}^0\Xi^{\ \alpha\gamma\ (m_s,m_s')}_{l_am_{l_a}l_cm_{l_c}}(\Omega)\ \left({A^{(\tau)}_{(m_s,m_s')}(\mathbf{r})
-\ x_0\ A^{0}_{(m_s,m_s')}(\mathbf{r})}\right)
    \end{split}
\end{align}
\begin{equation}
\label{eqn:HFDDFieldFinalPN}
\Gamma_{\alpha\gamma}^{(\tau_a\tau_c)} = t_3\int d\mathbf{r}\ R_{ac} \rho^\xi(\mathbf{r}) \left\lbrace{
(-x_0)\ \mathcal{Y}_{\alpha\gamma}\rho^{(\tau_a\tau_c)}(\mathbf{r})\ -\ \sum_{m_sm'_s}{}^0\Xi^{\ \alpha\gamma\ (m_s,m_s')}_{l_am_{l_a}l_cm_{l_c}}(\Omega)\  A^{(\tau_a\tau_c)}_{(m_s,m_s')}(\mathbf{r})
}\right\rbrace
\end{equation}
\begin{equation}
\Delta_{\alpha\beta}^{(\tau)} =
t_3(1-x_0) \int d\mathbf{r}\ R_{ab}\ \rho^\xi(\mathbf{r})\sum_{m_s,m_s'}
{}^2\Xi^{\ \alpha\beta\ (m_s,m_s')}_{l_am_{l_a}l_bm_{l_b}}(\Omega)
\left({{}^1B^{(\tau)}_{(m_sm_s')}(\mathbf{r}) - {}^1B^{(\tau)}_{(m_s'm_s)}(\mathbf{r})}\right)
\label{eqn:pairingFieldDD}
\end{equation}
\begin{equation}
\Delta_{\alpha\beta}^{(\tau_a\tau_b)}  =
t_3\int d\mathbf{r}\ R_{ab}\ \rho^\xi(\mathbf{r}) \sum_{m_sm'_s} 
{}^2\Xi^{\ \alpha\beta\ (m_s,m_s')}_{l_am_{l_a}l_bm_{l_b}}
\left({{}^1B^{(\tau_a\tau_b)}_{(m_sm'_s)}(\mathbf{r}) + x_0{}^1B^{(\tau_a\tau_b)}_{(m'_sm_s)}(\mathbf{r}) - {}^1B^{(\tau_b\tau_a)}_{(m'_sm_s)}(\mathbf{r})
- x_0{}^1B^{(\tau_b\tau_a)}_{(m_sm'_s)}(\mathbf{r})
}\right)
\end{equation}
%
%
%REARRANGEMENT
%
The variational method needs the rearrangement field and its \textit{pn} extensions. The Dirac delta matrix element $d_{\varepsilon\eta}(\mathbf{r})$ cancel the $\partial\Gamma_{\varepsilon\eta}^{pn}$ submatrices, however, the \textit{pn} parts contributes to the rearrangement in the \textit{pp-nn} terms:
\begin{equation}
\label{eqn:rearrangementFieldLikeParticles}
\partial\Gamma_{\varepsilon\eta}=
\alpha \sum_{\alpha\beta\gamma\delta}\langle{\alpha\beta|d_{\varepsilon\eta}(\mathbf{r})\rho^{\xi-1}(\mathbf{r})|\overline{\gamma\delta}}\rangle \left({2\rho_{\gamma\alpha}\rho_{\delta\beta} - \kappa_{\alpha\beta}^*\kappa_{\gamma\delta}}\right)
=\alpha t_3 \int d\mathbf{r}\ d_{\varepsilon\eta}(\mathbf{r})\ \rho^{\xi-1}(\mathbf{r})
\biggl({2\mathcal{D}ir(\mathbf{r})
+
\mathcal{P}air(\mathbf{r})
}\biggl)
\end{equation}
In the numerical evaluation of Eq.~\eqref{eqn:rearrangementFieldLikeParticles}, floating-point overflows in the low-density tail ($\rho \to 0$) are avoided by imposing a small density threshold, below which the rearrangement integrand is explicitly set to zero, consistent with its analytical limit.
\begin{equation}
\mathcal{D}ir(\mathbf{r})=
(\rho^0)^2-x_0\left({(\rho^p)^2+(\rho^n)^2}\right) - \sum_{m_sm_s'}\left({A^p_{(m_sm_s')}A^p_{(m_s'm_s)} + A^n_{(m_sm_s')}A^n_{(m_s'm_s)} - x_0A^0_{(m_sm_s')}A^0_{(m_s'm_s)}}\right)
\end{equation}
\begin{equation}
\mathcal{P}air(\mathbf{r})
=
(1-x_0)\sum_{m_sm_s'}
{}^2B^{(\tau)}_{(m_sm_s')}
\left({{}^1B^{(\tau)}_{(m_sm_s')} - {}^1B^{(\tau)}_{(m_s'm_s)}}\right)
\end{equation} 
\begin{equation}
\mathcal{D}ir^{(pn)}+\mathcal{D}ir^{(np)}= -x_0\ \rho^{pn}(\mathbf{r})\rho^{np}(\mathbf{r})\ 
- \sum_{m_sm_s'}\left({A^{pn}_{(m_s'm_s)}A^{np}_{(m_sm_s')}\ +\ A^{np}_{(m_s'm_s)}A^{pn}_{(m_sm_s')}}\right)
\label{eqn:ReaFieldPNRhoParts}
\end{equation}
\begin{equation}
\mathcal{P}air^{(\tau_a\tau_b)}(\mathbf{r})=\sum_{m_sm_s'} {}^2 B^{\tau_a\tau_b}_{(m_sm_s')}
\left\lbrace{
{}^1 B^{\tau_a\tau_b}_{(m_sm_s')} + x_0 {}^1 B^{\tau_a\tau_b}_{(m_s'm_s)} - \left({{}^1 B^{\tau_b\tau_a}_{(m_s'm_s)} + x_0 {}^1 B^{\tau_b\tau_a}_{(m_sm_s')}}\right)
}\right\rbrace
\end{equation}
\end{widetext}

From the antisymmetry property of $(\kappa^{(pn)})^T = -\kappa^{(np)}$, the bulk fields $B$ follow the relation
\begin{equation}
    \label{eq:B12ms_difft_definitions}
    {}^{(1,2)}B^{(\tau_a\tau_b)}_{(m_s,m_s')}(\mathbf{r}) = -{}^{(1,2)}B^{(\tau_b\tau_a)}_{(m_s',m_s)}(\mathbf{r})
\end{equation}
From this property, we can omit the lower-triangular calculation of the pairing field, $\Delta^{(\tau_b\tau_a)}_{\beta\alpha}=-\Delta_{\alpha\beta}^{(\tau_a\tau_b)}$ (changing the spin index $m_s\leftrightarrow m_s'$). From Eq.~\eqref{eqn:pairingFieldDD}, we see that the $\tau_a=\tau_b$ case follows the same rule.

For efficient computation, these fields are evaluated in the code using the associated Laguerre quadrature for the radial part ($\int x^{1/2}e^{-x} f(x)dx$ with variable $r/b=\sqrt{x/(\alpha+2)}$ with weigths and roots $\{x_i,w^{Lag}_{i_{x}}\}$) and the Lebedev quadrature \cite{LebedevQuadr} for the angular part, with roots and weights $\{\mathbf{\Omega}_{i_{\mathcal{O}}},w_{i_\mathcal{O}}^{Leb}\}$ given in symmetric sets of order $\mathcal{O}$. The matrix elements and the field evaluation follow this quadrature scheme
\begin{align}
    \begin{split}
\overline{v}^{DD}_{\alpha\beta\gamma\delta}=
\ &
\dfrac{t_3\  b^3}{2(\alpha+2)^{3/2}} 
\sum_{i_x} w^{Lag}_{i_x}\ {\overline{R}}_{abcd}\left({r(x_i)}\right) \times
\\
&\ 4\pi\sum_{i_\mathcal{O}} w^{Leb}_{i_\mathcal{O}}\ \overline{\rho}^{\ \alpha}(r(x_i), \mathbf{\Omega}_{i_\mathcal{O}})
\mathcal{A}_{\alpha\beta\gamma\delta}(\mathbf{\Omega}_{i_\mathcal{O}})
\\
\mathcal{A}_{\alpha\beta\gamma\delta}=&
\mathcal{Y}_{\alpha\gamma}\mathcal{Y}_{\beta\delta}(\mathbf{\Omega}_{i_\mathcal{O}})X_{abcd} - 
\mathcal{Y}_{\alpha\delta}\mathcal{Y}_{\beta\gamma}(\mathbf{\Omega}_{i_\mathcal{O}})X_{abdc}
    \end{split}
\end{align}
with $\rho(\mathbf{r}) = \overline{\rho}(\mathbf{r})e^{-(r/b)^2}$.
\\

\section{Benchmarking of the isospin-separated code.}\label{Benchmark}
\begin{figure*}
\centering
\includegraphics[width=\textwidth]{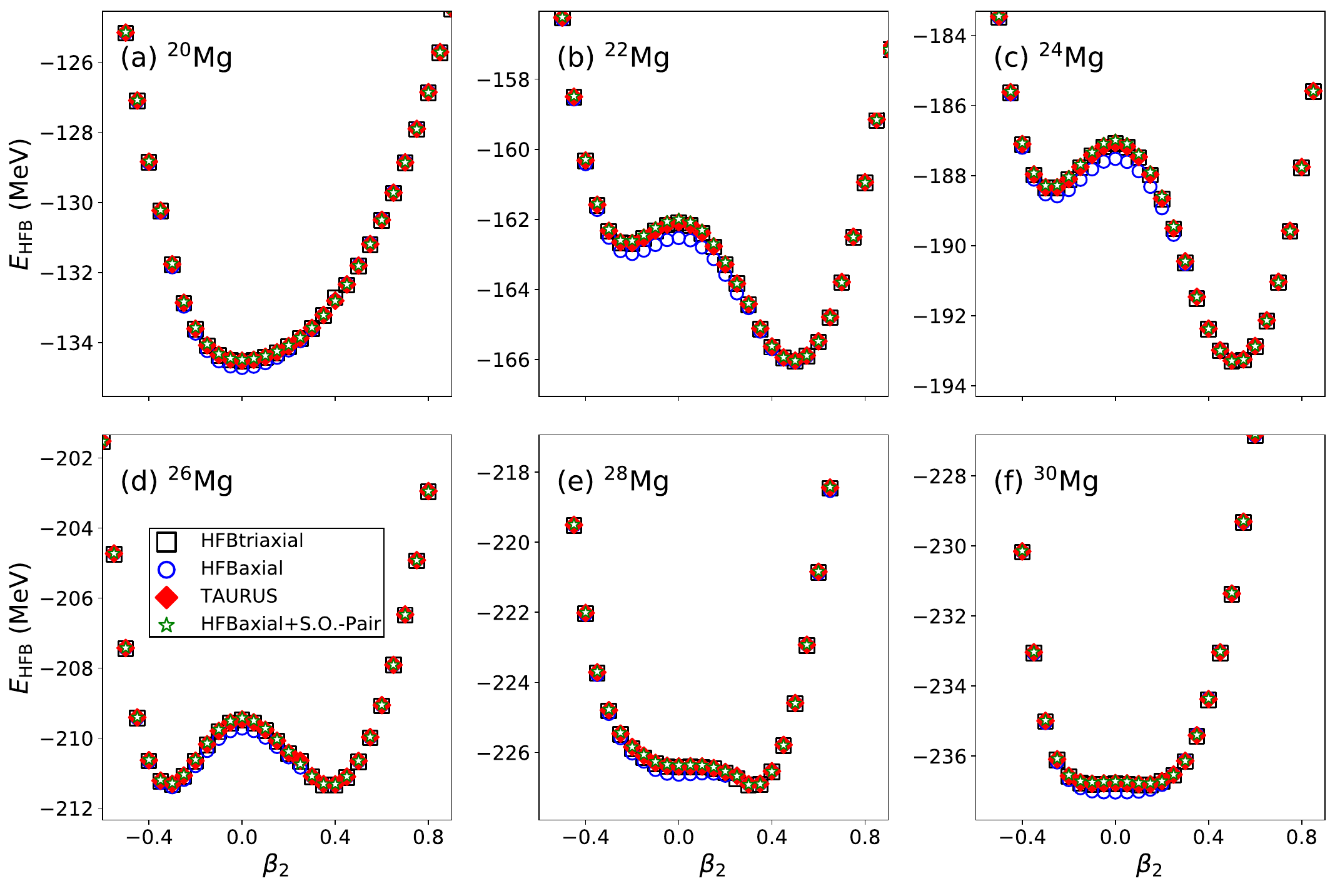}
\caption{Benchmark results for the implemented D1S interaction in \texttt{TAURUS} in the even-even Magnesium chain.}
\label{fig:benchmark_b20axial}
\end{figure*}
In this work, the \texttt{TAURUS} code has been extended for the first time to allow calculations with energy density functionals such as Gogny. This code employs a spherical harmonic-oscillator basis written in spherical coordinates as its working basis and is able to break all symmetries of the system (except complex HFB transformations), including proton-neutron mixing. The HFB equations are solved using the gradient method. The calculation of the energy and the gradient is performed directly from the two-body matrix elements of the interaction. Since this is a novel implementation, two highly non-trivial tests are discussed in this Appendix in order to verify in certain limiting cases that the implementation is correct. To this end, two independent and widely used benchmark codes that solve the HFB equations with the Gogny interaction are employed, namely \texttt{HFBaxial}~\cite{Robledo11a} and \texttt{HFBtriaxial}~\cite{EgidoRobledoTriaxial,Anguiano01a}. These codes use different working bases (axial and triaxial bases, respectively), different implementations of the gradient method (standard gradient with second-order corrections and conjugate gradient, respectively), and allow for the breaking of different self-consistent symmetries (axial symmetry with parity breaking, and triaxial states without parity breaking, respectively). In addition, the \texttt{HFBtriaxial} code includes all terms of the functional, whereas \texttt{HFBaxial} neglects the spin-orbit pairing term. Since these codes are highly optimized for the Gogny functional, the energy and the gradient are computed from intermediate fields derived in the most analytical way possible from the two-body matrix elements, the density matrix, and the pairing tensor. Hence, the two-body matrix elements are neither stored nor used explicitly in these codes.
\begin{table}{}
\begin{tabular}{l|rrr|}
\cline{2-4}
         \textbf{\texttt{HFBaxial}} & \multicolumn{1}{c|}{protons} & \multicolumn{1}{c|}{neutrons} & Total \\  \hline
\multicolumn{1}{|c|}{$K$} & 79.635866  & \multicolumn{1}{r|}{142.442577} & 222.078444  \\
\multicolumn{1}{|c|}{$-E_{HF}$} & 141.490046 & \multicolumn{1}{r|}{174.991229} & 316.481276 \\
\multicolumn{1}{|c|}{$-E_{pair}$} & 3.536002  & \multicolumn{1}{r|}{3.853228} & 7.389229 \\ \hline
\multicolumn{1}{|c|}{$-E_{HFB}$} &  65.390181 & \multicolumn{1}{r|}{36.401879} & 101.792061 \\ \hline
\multicolumn{1}{|c|}{$\sqrt{\langle{r^2}\rangle}$} & 2.647541 & \multicolumn{1}{r|}{2.957228} & 2.845048 \\ \hline
\multicolumn{1}{|c|}{$\sigma$} & 2.193301 & \multicolumn{1}{r|}{2.490340} & 4.683641 \\ \hline
\multicolumn{1}{|c|}{$Tr(\rho\ \partial\Gamma^{DD})$} & 39.894516 & \multicolumn{1}{r|}{56.518540} & 96.413057 \\ \hline
\end{tabular}
\\
\begin{tabular}{l|rrr|}
\cline{2-4}
    \textbf{\texttt{TAURUS} } & \multicolumn{1}{c|}{protons} & \multicolumn{1}{c|}{neutrons} & Total \\  \hline
\multicolumn{1}{|c|}{$K$} & 79.635860  & \multicolumn{1}{r|}{142.442564} & 222.078424  \\
\multicolumn{1}{|c|}{$-E_{HF}$} & 141.489829 & \multicolumn{1}{r|}{174.991013} & 316.480842 \\
\multicolumn{1}{|c|}{$-E_{pair}$} & 3.535994  & \multicolumn{1}{r|}{3.853209} & 7.389203 \\ \hline
\multicolumn{1}{|c|}{$-E_{HFB}$} &  65.389964 & \multicolumn{1}{r|}{36.401658} & 101.791622 \\ \hline
\multicolumn{1}{|c|}{$\sqrt{\langle{r^2}\rangle}$} & 2.647540 & \multicolumn{1}{r|}{2.957227} & 2.845048 \\ \hline
\multicolumn{1}{|c|}{$\sigma$} & 2.1932992 & \multicolumn{1}{r|}{2.4903325} & 4.683632 \\ \hline
\multicolumn{1}{|c|}{$Tr(\rho\ \partial\Gamma^{DD})$} &   & \multicolumn{1}{r|}{} & 96.413204 \\ \hline
\end{tabular}
\caption{Energy and observable results for $^{16}\text{C}$, using an axial wave function obtained with the \texttt{HFBaxial} code with the Gogny D1S interaction.}
\label{tab:C16-table}
\end{table}
The first test concerns the calculation of the Gogny D1S energy and other expectation values, such as the root-mean-square radius, using \texttt{HFBaxial} and \texttt{TAURUS} with a wave function originally obtained with \texttt{HFBaxial}. In order to ensure that the same wave function is used in both codes, the $(U,V)$ matrices of the Bogoliubov transformation, originally expressed in the spherical harmonic-oscillator basis in cylindrical coordinates employed in \texttt{HFBaxial}, are transformed to the spherical basis used in \texttt{TAURUS}. Table~\ref{tab:C16-table} shows the results for a wave function of the nucleus $^{16}$C. Only minimal differences between the results obtained with the two codes are observed.

The second, more demanding test consists of calculating the HFB energy curves as a function of the quadrupole deformation $\beta_{2}$ with the three aforementioned programs for the even–even nuclei $^{20-30}$Mg. In all cases, five harmonic-oscillator shells and the same oscillator length are used. In this test, both the energy and the gradient must be correctly evaluated in order to converge to the same solution, provided that the same functional is minimized. Figure~\ref{fig:benchmark_b20axial} shows that this is indeed the case for the \texttt{TAURUS} and \texttt{HFBtriaxial} codes at all calculated points, and for the \texttt{HFBaxial} code at those points where the spin–orbit pairing energy is completely negligible. However, in regions close to spherical deformation, \texttt{TAURUS} and \texttt{HFBtriaxial} yield slightly higher energies than \texttt{HFBaxial}. This is due to the fact that this contribution is not included either in the functional being minimized or in the evaluated energy. Nevertheless, since the method is variational, when the wave functions converged with \texttt{HFBaxial} are used to compute the full energy, this energy lies above that obtained with \texttt{TAURUS} and \texttt{HFBtriaxial}. In any case, the differences arising from the inclusion or exclusion of the spin–orbit pairing term in the minimization are very small and, at least for the present calculations, neglecting this term in the \texttt{HFBaxial} code is justified.
\bibliographystyle{apsrev4-2}
\bibliography{biblio}
%%%%%%%%%%%%%%%%
\end{document}